\documentclass[aps,twocolumn,superscriptaddress,amsmath,showpacs,tightenlines]{revtex4}
\usepackage{graphicx}
\usepackage{epsfig}
\begin{document}
\title{Robustness of Charge-Qubit Cluster States to Double Quantum Point Contact Measurement}

\author{Tetsufumi Tanamoto}
\affiliation{Corporate R \& D center, Toshiba Corporation,
1 Komukai-Toshiba cho, Saiwai-ku, Kawasaki 212-8582, Japan}

\begin{abstract}
We theoretically investigate the robustness of cluster states in charge qubit system based on quantum dot (QD)
and double quantum point contact (DQPC). Trap state is modeled by an island structure in DQPC and represents 
a dynamical fluctuation. We found that the dynamical fluctuations affect the cluster states more than static fluctuation 
caused by QD size fluctuation.
\end{abstract}

\maketitle
\section{Introduction}
One-way quantum computing~\cite{Briegel1,Briegel2} 
is an important approach for quantum computation 
based on a series of one-qubit measurements starting 
from a cluster state of a qubit array. 
Cluster states are highly-entangled states involving 
all qubits and are typically generated from 
an Ising-like Hamiltonian, 
starting from an initial product state 
$|\Psi_0\rangle \equiv |\Psi(t=0)\rangle = \Pi_{i=1}^{N} |+\rangle_i\,$, where 
$|\pm \rangle_i=(|0\rangle_i\!\pm\!|1\rangle_i)/\sqrt{2}$. 
Here, $|0\rangle_i$ and $|1\rangle_i$ are the two states of the $i$-th qubit 
in an $N$-qubit system.  
In ref.~\cite{tana1}, we showed that cluster states in charge 
qubits\cite{Single,Fujisawa,Gorman,tana0,tana0a,Ahmed,Fossi,Hawrylak,Liu,Aguado,Brandes,Goan,Koiller,Gilad} can 
be created by applying a single gate bias pulse, 
right after preparing the initial product state (one-step generation method), 
and are more robust against nonuniformities 
among qubits than decoherence-free (DF) states~\cite{Zanardi,Lidar} 
under a noise environment generated by a quantum point contact (QPC) detector, 
which is a sensitive detector of electric charge distribution\cite{Field,Yuan}. 
However, trap sites are often unavoidable 
in solid-state qubits owing to their small fabrication size and
several experiments show that 
trap states significantly affect electric transport properties of nanoscale devices\cite{Sakamoto,Peters,Nakamura,Jung}.
Here, we model the trap site as the island (discrete energy state) between double QPCs (DQPC)   
and investigate robustness of cluster states in charge qubits 
measured by the DQPC detector~(Fig. 1). 
The charge qubits are based on quantum dots (QD), in which the position of the excess charge in a qubit 
affects the QPC current electrically, resulting in detection of charged state. 
Owing to this additional external degrees of freedom, this setup is considered to be a harsher and 
more realistic condition for qubits than that of ref~.\cite{tana1}.
Here, we calculate a time-dependent fidelity of four charge qubits 
with DQPC by solving density matrix (DM) equations.

This article is organized as follows. In \S 2, formulation 
of our model is presented. Section 3 is devoted to the numerical 
calculations for the cluster state and DF state. 
The conclusion is given in \S 4 .
The detailed coefficients used in the equations in the text are 
shown in Appendix.

\begin{figure}[h]
\begin{center}
\includegraphics[width=5cm]{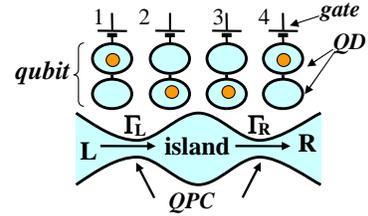}
\vspace*{4cm}
\caption{Four qubits that use double QD charge states are capacitively 
coupled to a double QPC detector with an island. One excess charge is injected into each qubit.
Single energy level is assumed in each QD and the island.
This figure shows $|1001\rangle$ state.
}
\label{QPC}
\end{center}
\end{figure}

\section{Density Matrix Equations}
Here we show the formulation of the density matrix equations for Fig.~\ref{QPC}. 
The Hamiltonian for the combined qubits and the QPC for Fig.~\ref{QPC}
is written as $H = H_{\rm qb}\!+\!H_{\rm qpc}\!+\!H_{\rm int}$.  
$H_{\rm qb}$ is a Hamiltonian for an
array of charge qubits with nearest-neighbor interactions, 
expressed by~\cite{tana1,tana0,tana0a}
\begin{equation}
H_{\rm qb}=\sum_i (\Omega_i \, \sigma_{ix}
+\epsilon_{i}\,\sigma_{iz})
+\sum_{i<j}J_{ij}\,\sigma_{iz}\,\sigma_{jz}
, \label{Hcq}
\end{equation}
where $\sigma_{ix}$ and $\sigma_{iz}$ are Pauli matrices 
for the $i$-th qubit expressed by 
$\sigma_{ix}=|0\rangle_i \langle 1| +|1\rangle_i \langle 0|$ 
and  $\sigma_{iz}=|0\rangle_i \langle 0| -|1\rangle_i \langle 1|$. 
$|0\rangle$ and $|1\rangle$ refer
to the two single-qubit states in which the excess charge is localized in the
upper and lower dot, respectively.  
$\Omega_i$ is the inter-QD tunnel
coupling between two coupled QDs.
$\epsilon_i$ is the charging energy and corresponds to the energy difference 
between $|0\rangle_i$ and $|1\rangle_i$ controlled by gate voltage. 
The coupling constants $J_{ij}$ are derived from the capacitance 
couplings~\cite{tana_m}.

The Hamiltonian for the two serially coupled QPCs $H_{\rm qpc}$ is described by
\begin{eqnarray}
H_{\rm qpc} \!\!&=&\!\!  
\sum_{\alpha=L,R \atop s\!=\!\uparrow,\downarrow}\! 
\sum_{\ i_\alpha} \left[ E_{i_\alpha} c_{i_\alpha s}^\dagger
c_{i_\alpha s} \!+\! V_{i_\alpha s} (c_{i_\alpha s}^\dagger d_{s} +
d_{s}^\dagger c_{i_\alpha s} )\right]
\nonumber \\
&+& \sum_{s\!=\!\uparrow,\downarrow} E_d d_{s}^\dagger d_{s} 
+ U d_{\uparrow}^\dagger d_{\uparrow} d_{\downarrow}^\dagger d_{\downarrow}\,.
\label{eqn:H_qpc}
\end{eqnarray}
\normalsize
Here, $c_{i_{L}s}$ ($c_{i_{R}s}$) is the annihilation operator of an electron
in the $i_L$th ($i_R$th) level $[i_L (i_R)=1,2,3,...]$ of the left (right)
electrode, $d_{s}$ is the electron annihilation operator of the island
between the QPCs, $E_{i_{L}s}$ ($E_{i_{R}s}$) is the energy level of electrons 
in the left (right) electrode, and $E_d$ is that of the island.  Here, we
assume only one energy level on the island between the two QPCs, with spin
degeneracy.  $V_{i_Ls}$ ($V_{i_Rs}$) is the tunneling strength of electrons
between the left (right) electrode state $i_Ls$ ($i_Rs$) and the island state.  
$U$ is the on-site Coulomb energy of double occupancy in the island. 

$H_{\rm int}$ is the capacitive interaction between the qubits and the QPC,
which induces {\it dephasing} between different eigenstates of $\sigma_{iz}$
\cite{Zanardi,Lidar}.
Most importantly, it takes into account the fact 
that localized charge near the QPC
increases the energy of the system electrostatically, thus affecting the
tunnel coupling between the left and right electrodes: 
\begin{eqnarray}
H_{\rm int} &=& 
\sum_{i_L,s} \left[ \sum_{i=1}^2 \delta V_{i_L,is}\sigma_{iz} \right]
( c_{i_L s}^\dagger d_s + d_s^\dagger c_{i_L s} ) \nonumber \\
&+& 
\sum_{i_R,s} \left[ \sum_{i=3}^4 \delta V_{i_R, is}\sigma_{iz} \right]
( c_{i_R s}^\dagger d_s + d_s^\dagger c_{i_R s} ) 
\end{eqnarray}
where $\delta V_{i_\alpha is}$ ($\alpha=L,R$) 
is an effective change of the tunneling strength
between the electrodes and QPC island.  
Hereafter, we neglect the spin dependence of $V_{i_\alpha}$ and 
$\delta V_{i_\alpha,i}$.
We assume that the tunneling strength of electrons weakly depends 
on the energy $V_{i_\alpha i}= V_{i\alpha}(E_{i_\alpha})$ 
and electrodes are degenerate up to the Fermi
surface $\mu_\alpha$.  
Then qubit states influence the QPC tunneling rate $\Gamma_L$ and $\Gamma_R$ 
by $\Gamma_L^{-1}\!=\!\Gamma_{1}^{-1}\!+\!\Gamma_{2}^{-1}$
and $\Gamma_R^{-1}\!=\!\Gamma_{3}^{-1}\!+\!\Gamma_{4}^{-1}$ 
through $\Gamma_{i}^{(\pm)} \equiv
2\pi \wp_{\alpha} (\mu_\alpha) |V_{i\alpha}^{(\pm)} (\mu_\alpha) |^2$ and
$\Gamma_{i}^{(\pm)'} \!
\equiv\! 2\pi \wp_{\alpha} (\mu_\alpha\!+\!U) 
|V_{i\alpha}^{(\pm)} (\mu_\alpha\!+\!U)|^2$, 
depending on the qubit state 
$\sigma_{iz}=\pm 1$ 
($V_{i\alpha}^{(\pm)}(\mu_\alpha) =
V_{i\alpha}\!(\mu_\alpha) \pm \delta V_{i\alpha}(\mu_\alpha)$ 
and $\wp_{\alpha} (\mu_\alpha)$ is the density of states of the electrodes
($\alpha\!=\!L,R$) and
$\Gamma_i'$ is a tunneling rate when the island lies in the ``c" state). 
The values of $\Gamma_{i}^{(\pm)}$s are determined 
by the geometrical structure
of the system and depend on the distance between qubits and the QPC
as $\Gamma_{i}^{(\pm)}\!=\Gamma_{i0}\!\pm\Delta \Gamma_{i}
=\Gamma_0 (1\pm\zeta)$ 
with the measurement strength $\zeta\equiv\Delta\Gamma/\Gamma_0$~\cite{tana2}.

The DM equations of four qubits and the DQPC detector at zero temperature 
of Fig.~\ref{QPC} are derived similarly to ref.~\cite{tana2,Li} by
\begin{eqnarray}
& &\frac{d \rho_{z_1z_2}^{a}}{dt}\!=\!(\!i[J_{z_2}\!-\!J_{z_1}\!]
-\![\Gamma_L^{z_1}\!+\!\Gamma_L^{z_2} ])\rho_{z_1z_2}^{a}
\nonumber \\
\!&\!-\!&\! i\!\sum_{j=1}^{N}\Omega_j 
(\rho_{g_j(z_1),z_2}^{a}\!-\!\rho_{z_1,g_j(z_2)}^{a})
\!+\! \sqrt{\Gamma_R^{z_1}\Gamma_R^{z_2}} 
(\rho_{z_1z_2}^{b\uparrow}+\rho_{z_1z_2}^{b\downarrow}),
\nonumber \\
& &\frac{d \rho_{z_1z_2}^{b_s}}{dt}\!=\!\left(\!i[J_{z_2}\!-\!J_{z_1}]\!
-\!\frac{\Gamma_L^{z_1'} \!+\!\Gamma_L^{z_2'} 
\!+\!\Gamma_R^{z_1}\!+\!\Gamma_R^{z_2}}{2}\right)
\rho_{z_1z_2}^{b_s}
\nonumber \\
\!&\!-\!&\! i\!\sum_{j=1}^N \Omega_j 
(\rho_{\!g_j(z_1),z_2}^{b_s}\!\!-\!\!\rho_{\!z_1,g_j(z_2)}^{b_s})
\!+\! \!
\sqrt{\Gamma_L^{z_1}\Gamma_L^{z_2}} \!\rho_{z_1\!z_2}^{a}\!
\!+\! \!\sqrt{\Gamma_R^{z_1'}\Gamma_R^{z_2'}}\! \rho_{z_1\!z_2}^{c},
\nonumber \\
& &\frac{d \rho_{z_1z_2}^{c}}{dt}
\!=\!(\!i[J_{z_2}\!-\!J_{z_1}]
\!-\![\Gamma_R^{z_1'}+\Gamma_R^{z_2'}] )
\rho_{z_1z_2}^{c}
\nonumber \\
\!&\!-\!&\! i\!\sum_{j=1}^N \Omega_j 
(\rho_{g_j(z_1),z_2}^{c}\!-\!\rho_{z_1,g_j	(z_2)}^{c})
\!+\! \sqrt{\Gamma_L^{z_1'}\Gamma_L^{z_2'}}
(\rho_{z_1z_2}^{b\uparrow}+\rho_{z_1z_2}^{b\downarrow}),
\nonumber\\ \!\!\!\!\!\!\!\!\!\!\!\!\!\!\!\!\!\!\!\!\!
\label{eqn:dm}
\end{eqnarray}
where $z_1,z_2$ indicate qubit states such as $0000,0001,...,1111$ and, 
$\rho_{z_1z_2}^{a}$, $\rho_{z_1z_2}^{b_s}$ and 
$\rho_{z_1z_2}^{c}$ are density matrix elements 
when no electron (``a"), one electron (``b") and two electrons (``c") 
exist in the QPC island, respectively. 
$J_{0000} = \sum_i^4 \epsilon_i+J_{12}+J_{23}$, 
$J_{0001} = \sum_i^3 \epsilon_i-\epsilon_4+J_{12}-J_{23}$, 
...,
$J_{1111} =-\sum_i^4\epsilon_i+J_{12}+J_{23}$.
$g_j(z_i)$ is introduced for the sake of notational 
convenience and determined by the relative positions between qubit
states (see Appendix and ref.~\cite{tana2}).
In the present case, there are 768 equations to be solved. 

Time-dependent {\it fidelity}
$F(t)\!\equiv\! {\rm Tr}[\hat{\rho}(0) \hat{\rho}(t)]$
is calculated by tracing over the elements 
of the reduced DM obtained from eq.~(\ref{eqn:dm}).

We use these DM equations to describe four 
qubits, and compare fidelity of a cluster state 
\begin{eqnarray}
|\Psi\rangle_{\rm CS}&=&\frac{1}{2} \left(|+\rangle_1 |0\rangle_2
|+\rangle_3|0\rangle_4 \!+\!|+\rangle_1 |0\rangle_2
|-\rangle_3|1\rangle_4 \right. \nonumber \\
&+&\left. |-\rangle_1 |1\rangle_2
|-\rangle_3|0\rangle_4 \!+\!|-\rangle_1 |1\rangle_2
|+\rangle_3|1\rangle_4 \right)
\end{eqnarray}
with that of a DF state 
\begin{equation}
|\Psi\rangle_{\rm DF}=\frac{1}{2} (|1100\rangle
-|1001\rangle 
-|0110\rangle
+|0011\rangle).
\end{equation} 

\section{Cluster States in Charge Qubits}
In one-way quantum computing, quantum computation 
is carried out by measuring a series of qubits in cluster states.
Cluster states are directly generated if we can 
prepare an Ising-like Hamiltonian 
$H_{\rm cs}=(g/4)\sum_{i<j} (1-\sigma_{iz})(1-\sigma_{jz})$, 
starting from the initial state $|\Psi_0\rangle$.
A unitary evolution $U_{\rm cs}(t)=\exp (-itH_{\rm cs})$ (we use $\hbar=1$) 
at $gt=(2n_I+1)\pi$ with an integer $n_I$ transforms $|\Psi_0\rangle$ into 
cluster states. In the case of charge qubit in eq.~(\ref{Hcq}), the $\sigma_{ix}$ term needs to be
switched off during the creation of the cluster state and
then switched on when measurements are carried out.
To realize this controllability, additional gates are required. However, these additional gates make the qubit system 
complicated, ending by producing decoherence and crosstalk between qubits themselves and between
qubits and the environment. In addition, for some qubit
systems, once the qubit array is made, $\Omega_i$ and $J_{ij}$ are fixed,
and only $\epsilon_i$ is controllable via the gate voltage bias (we call
these ``simple-design qubits''). 
In ref.~\cite{tana1}, we proposed how to effectively generate 
cluster states in such simple-design qubits (``one-step generation 
method").
In this method, cluster states are obtained by applying a gate bias voltage $\epsilon_i$ 
for the $i$-th qubit, expressed as
\begin{equation}
\epsilon_i = \epsilon_i^{\rm cs}=
\sqrt{E_{i}^2- \Omega_i^2},    
\end{equation}
at time $t_{\rm cs}= \pi(8n_{J}+1)/(4J)$. 
Here, $E_i\!=(\epsilon_i+\Omega_i^2/\epsilon_i)\cos\alpha_i$ 
with $\tan \alpha_i=\Omega_i /\epsilon_i  \ll 1$, 
and we also need a relation $J(8n_{E}-\bar{n}_i)
/(8n_{J}+1)=E_i$ with an arbitrary integer $n_J$  and 
the number of the nearest qubits $\bar{n}_i$.
As long as these equations are held, we can choose any 
integers $n_J$ and $\bar{n}_i$. 

In the present one-dimensional qubit array, 
we have $E_1=E_4=(8 n_E-1)J$ and $E_2=E_3=(8 n_E-2)J$. 
Here, we calculate two examples: 
case $\{1\}$ is $\Omega=2J$ ($n_E=1$) 
and case $\{2\}$ is $\Omega=4J$ ($n_E=1$). 
In case $\{1\}$, 
we have 
$\epsilon_1^{\rm cs}=\epsilon_4^{\rm cs}\approx 6.7J$, 
$\epsilon_2^{\rm cs}=\epsilon_3^{\rm cs}\approx 5.7J$.
In case $\{2\}$
we have 
$\epsilon_1^{\rm cs}=\epsilon_4^{\rm cs}\approx 5.7J$,  
$\epsilon_2^{\rm cs}=\epsilon_3^{\rm cs}\approx 4.5J$.

\begin{figure}
\begin{center}
\includegraphics[width=5.5cm]{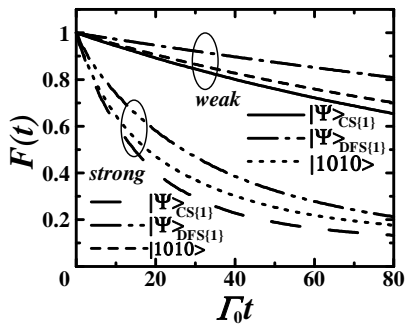}
\vspace*{4.3cm}
\caption{Time-dependent fidelity of four-qubit states 
for a cluster state, a DF state, and 
a product state $|1010\rangle$, in case $\{1\}$. 
$\Gamma_0=J$, $\Gamma'_{i}=\Gamma_{i}$.
The ``weak'' indicates a weak measurement case of 
$\zeta=0.2$ and the ``strong'' indicates a strong measurement 
case of $\zeta=0.6$.
}
\label{Fstruct}
\end{center}
\end{figure}

\section{Numerical Calculations}


Figure \ref{Fstruct} shows a time-dependent fidelity when the 
measurement strength is changed.  
It can be seen that when the measurement strength is sufficiently large 
($\zeta=0.6$), the fidelity of cluster states greatly degrades.
This result shows that the cluster state is sensitive to the existence of 
trap sites. 

\begin{figure}[h]
\begin{center}
\vspace*{0.3cm}
\includegraphics[width=8.5cm]{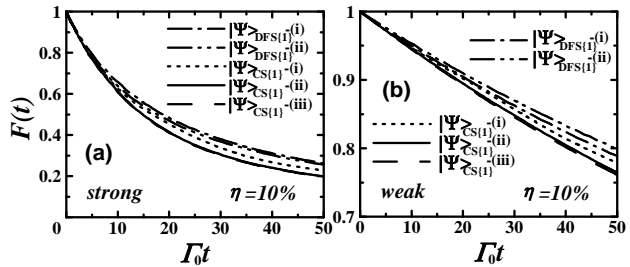}
\vspace*{3.5cm}
\caption{Time-dependent fidelities $F(t)$
of a cluster state and a DF state when 
there are nonuniformities in qubit parameter.
$\Omega=4J$ and $n_{E}=1$ (case $\{1\}$). 
$\Gamma_0=J$, $\Gamma'_{i}=\Gamma_{i}$, 
and $\zeta=0.6$. 
(a) Strong measurement case.
(b) Weak measurement case. Nonuniformity in the 
qubit parameters is introduced as
$\Omega_i\!=2J \!(1\!-\!\eta_i)$, $\epsilon_i\!=\epsilon+\!\eta_i
J$, and $\Gamma_i^{(\pm)}\!=\!(1\!-\!\eta_i)\Gamma^{(\pm)}$, with
$i$ indicating the $i$-th qubit. 
Here $\eta_i=0$ for all qubits besides $\eta_4=0.1$ (i),
$\eta_2=\eta_3=0.1$ (ii), and $\eta_4=0.1$ (iii). 
The fidelities of
$|\Psi\rangle_{\rm CS}$ for (i) and (ii) mostly overlap.
} 
\label{FID}
\end{center}
\end{figure}

Figures \ref{FID}~(a) and \ref{FID}~(b) show the combined effects of the island structure 
and the nonuniformities among qubits 
when parameters $\Omega_i$, $\epsilon_i$ and $\Gamma_i$ 
fluctuate by 10\%. 
The smaller sizes of QDs are preferable for qubits based on 
QDs\cite{tana0,tana0a,tana_m}. If we assume that a diameter of each QD is 5~nm\cite{Tiwari},
10\% size fluctuation corresponds to $\pm 5\AA$ size deviation from 5~nm.
In both the strong measurement case [(a) $\zeta=0.6$] 
and the weak measurement case [(b) $\zeta=0.2$],  the
difference between the cluster state and the DF state is small. 
This is in contrast to the case without local island 
structure in QPC discussed in ref.~\cite{tana1}, and shows that 
the island structure (trap site) imposes a larger decoherence environment 
and would be a major origin of degradation for the entangled states.

\begin{figure}[h]
\begin{center}
\includegraphics[width=6cm]{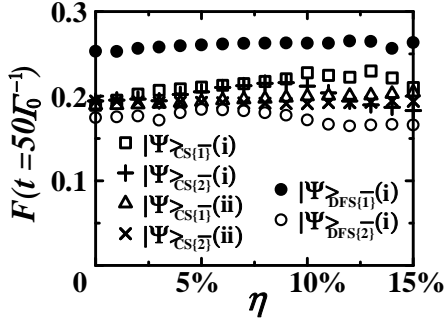}
\vspace*{4cm}
\caption{Fidelities of four-qubit cluster state and DF state 
at $t\!=\!50\Gamma_0^{-1}$ 
as a function of nonuniformity $\eta$.
Parameters are the same as those in Fig.~\ref{FID}.
}
\label{eta}
\end{center}
\end{figure}

In order to examine the effect of nonuniformity between qubits in more detail, 
we calculate a fidelity at time $t=50\Gamma_0$ as a 
function of nonuniformity of qubit parameter $\eta$ (Fig.~\ref{eta}).
We can see that the difference is slight for various 
types of nonuniformities when $\eta$ is sufficiently small. 
We can also see that the fidelity of the larger gate bias case $\{1\}$ 
is degraded slightly more than that of the smaller gate bias case $\{2\}$. 
Note that this difference is much smaller in cluster states 
than in DF states, 
which is a preferable result for one-way computing in solid-state 
qubits. 
From Figs.~\ref{FID} and \ref{eta}, we can confirm that 
the cluster state is robust to nonuniformity, 
whereas the cluster state is weak against local electronic state (trap state), 
which is similar to the DF state.


\section{Conclusions}
We investigated the effect of local electronic fluctuations (trap state) 
in addition to nonuniformities, as a model of a decoherence mechanism in cluster states.  
We found that the island (trap site) affects the fidelity of cluster 
states significantly. 
Because the local electronic state provides {\it dynamical} fluctuation whereas 
the nonuniformities provide {\it static} fluctuation, we can say that 
the cluster states are sensitive to {\it dynamical} fluctuations.
Experimentally, it is difficult to reduce the number of trap sites. Thus, 
more detailed understanding of effects of trap states will be required in the future.

\acknowledgements
The author thanks A. Nishiyama, J. Koga, S. Fujita, F. Nori, Y. X. Liu, and X. Hu for 
useful discussions.

\appendix
\section{Coefficients in DM Equations}
Here, we show all coefficients used in eq.~(\ref{eqn:dm}). 
For legibility, we denote $|00 \rangle$, $|01\rangle$, $|10 \rangle$, and $|11\rangle$ 
as $|A\rangle$, $|B\rangle$, $|C\rangle$, $|D\rangle$, respectively. 
Four-qubit states are written by $|AA\rangle$, $|AB\rangle$,
..,$|DD\rangle$. 
Then,  $J_{z_1} (z_1=AA,...,DD) $ are expressed as
{\small
\begin{eqnarray}
J_{AA}&=& \epsilon_1\!+\epsilon_2\!+\epsilon_3\!+\epsilon_4
 \!+J_{12}\!+J_{23}\!+J_{34},
\nonumber \\
J_{AB}&=& \epsilon_1\!+\epsilon_2\!+\epsilon_3\!-\epsilon_4
 \!+J_{12}\!+J_{23}\!-J_{34}, 
 \nonumber \\
J_{AC}&=& \epsilon_1\!+\epsilon_2\!-\epsilon_3\!+\epsilon_4
 \!+J_{12}\!+J_{23}\!-J_{34},
\nonumber \\
J_{AD}&=& \epsilon_1\!+\epsilon_2\!-\epsilon_3\!-\epsilon_4
 \!+J_{12}\!-J_{23}\!+J_{34},
 \nonumber \\
J_{BA}&=& \epsilon_1\!-\epsilon_2\!+\epsilon_3\!+\epsilon_4
 \!-J_{12}\!-J_{23}\!+J_{34},
\nonumber \\
J_{BB}&=& \epsilon_1\!-\epsilon_2\!+\epsilon_3\!-\epsilon_4
 \!-J_{12}\!-J_{23}\!-J_{34},
 \nonumber \\
J_{BC}&=& \epsilon_1\!-\epsilon_2\!-\epsilon_3\!+\epsilon_4
 \!-J_{12}\!+J_{23}\!-J_{34},
\nonumber \\
J_{BD}&=& \epsilon_1\!-\epsilon_2\!-\epsilon_3\!-\epsilon_4
 \!-J_{12}\!+J_{23}\!+J_{34},
 \nonumber \\
J_{CA}&=&-\epsilon_1\!+\epsilon_2\!+\epsilon_3\!+\epsilon_4
 \!-J_{12}\!+J_{23}\!+J_{34},
\nonumber \\
J_{CB}&=&-\epsilon_1\!+\epsilon_2\!+\epsilon_3\!-\epsilon_4
 \!-J_{12}\!+J_{23}\!-J_{34},
 \nonumber \\
J_{CC}&=&-\epsilon_1\!+\epsilon_2\!-\epsilon_3\!+\epsilon_4
 \!-J_{12}\!-J_{23}\!-J_{34},
\nonumber \\
J_{CD}&=&-\epsilon_1\!+\epsilon_2\!-\epsilon_3\!-\epsilon_4
 \!-J_{12}\!-J_{23}\!+J_{34},
 \nonumber \\
J_{DA}&=&-\epsilon_1\!-\epsilon_2\!+\epsilon_3\!+\epsilon_4
 \!+J_{12}\!-J_{23}\!+J_{34},
\nonumber \\
J_{DB}&=&-\epsilon_1\!-\epsilon_2\!+\epsilon_3\!-\epsilon_4
 \!+J_{12}\!-J_{23}\!-J_{34},
 \nonumber \\
J_{DC}&=&-\epsilon_1\!-\epsilon_2\!-\epsilon_3\!+\epsilon_4
 \!+J_{12}\!+J_{23}\!-J_{34},
\nonumber \\
J_{DD}&=&-\epsilon_1\!-\epsilon_2\!-\epsilon_3\!-\epsilon_4
 \!+J_{12}\!+J_{23}\!+J_{34}.
\end{eqnarray}
}
$g_i(z_1) (i=1,..,4, z_1=AA,...,DD)$ are given by
{\small
$$
\begin{array}{llll}
g_1(AA)\!=\!CA,&g_2(AA)\!=\!BA,&g_3(AA)\!=\!AC,&g_4(AA)\!=\!AB,  \\
g_1(AB)\!=\!CB,&g_2(AB)\!=\!BB,&g_3(AB)\!=\!AD,&g_4(AB)\!=\!AA,  \\
g_1(AC)\!=\!CC,&g_2(AC)\!=\!BC,&g_3(AC)\!=\!AA,&g_4(AC)\!=\!AD,  \\
g_1(AD)\!=\!CD,&g_2(AD)\!=\!BD,&g_3(AD)\!=\!AB,&g_4(AD)\!=\!AC,  \\
g_1(BA)\!=\!DA,&g_2(BA)\!=\!AA,&g_3(BA)\!=\!BC,&g_4(BA)\!=\!BB,  \\
g_1(BB)\!=\!DB,&g_2(BB)\!=\!AB,&g_3(BB)\!=\!BD,&g_4(BB)\!=\!BA,  \\
g_1(BC)\!=\!DC,&g_2(BC)\!=\!AC,&g_3(BC)\!=\!BA,&g_4(BC)\!=\!BD,  \\
g_1(BD)\!=\!DD,&g_2(BD)\!=\!AD,&g_3(BD)\!=\!BB,&g_4(BD)\!=\!BC,  \\
g_1(CA)\!=\!AA,&g_2(CA)\!=\!DA,&g_3(CA)\!=\!CC,&g_4(CA)\!=\!CB,  \\
g_1(CB)\!=\!AB,&g_2(CB)\!=\!DB,&g_3(CB)\!=\!CD,&g_4(CB)\!=\!CA,  \\
g_1(CC)\!=\!AC,&g_2(CC)\!=\!DC,&g_3(CC)\!=\!CA,&g_4(CC)\!=\!CD,  \\
g_1(CD)\!=\!AD,&g_2(CD)\!=\!DD,&g_3(CD)\!=\!CB,&g_4(CD)\!=\!CC,  \\
g_1(DA)\!=\!BA,&g_2(DA)\!=\!CA,&g_3(DA)\!=\!DC,&g_4(DA)\!=\!DB,  \\
g_1(DB)\!=\!BB,&g_2(DB)\!=\!CB,&g_3(DB)\!=\!DD,&g_4(DB)\!=\!DA,  \\
g_1(DC)\!=\!BC,&g_2(DC)\!=\!CC,&g_3(DC)\!=\!DA,&g_4(DC)\!=\!DD,  \\
g_1(DD)\!=\!BD,&g_2(DD)\!=\!CD,&g_3(DD)\!=\!DB,&g_4(DD)\!=\!DC.  
\end{array}
$$
}

\end{document}